\documentstyle[12pt]{article}
\begin{document}

\hfill{July 1995}

\vskip3cm

\centerline{\Large \bf On behaviour of critical lines near ferrimagnetic}
\vskip3pt
\centerline{\Large \bf phase in Higgs-Yukawa systems}
\vskip1.0truecm
\centerline{
Sergei V. Zenkin \footnote {Electronic address:
zenkin@inr.msk.su}}
\vskip0.2cm\centerline{Institute for Nuclear Research of the Russian
Academy of Sciences}
\centerline{60th October Anniversary Prospect 7a, 117312
Moscow, Russia}

\vskip 2cm

\begin{abstract}

We calculate within a mean-field approximation the slopes of the
critical lines near the point of appearing the ferrimagnetic phase
for the U(1) systems in the weak coupling regime. It is demonstrated
that the slope of one of the critical line is continuous, while
change of the slope of the other depends strongly on the number of
the fermion flavours. We also find that in the ferrimagnetic phase
near such a point the magnetization and the staggered magnetization
align orthogonally to each other. \\

\noindent PASC numbers: 11.15.Ha, 05.70.Fh, 14.80.Bn
\end{abstract}

\newpage

\section{Introduction}

In our previous paper devoted to the mean-field analysis of the phase
structure of the Higgs-Yukawa systems \cite{TZ} we have concluded
that intersection of the critical line separating the
ferromagnetic (FM) and paramagnetic (PM) phases with the line
separating the antiferromagnetic (AM) and PM phases always leads to
appearing the ferrimagnetic (FI) phase, and that the slopes of the
critical lines in the point of their intersection (denote it $B$
\footnote{Notation $A$ has been employed in \cite{TZ} for another
point.}) are continuous. These statements have been criticized
recently in \cite{Zar} where it was argued that the slopes are in
general discontinuous and that the FI phase may not appear at all in
the point $B$.

In this paper we demonstrate that in reference to the systems
considered in \cite{TZ} this criticism is justified only in part.
Namely, considering within the mean-field approximation the weak
coupling regime for the U(1) systems we show that in all the cases
when the intersection of the FM-PM and AM-PM critical lines occurs,
the FI phase does appear, and that the slope of the AM-PM--FI-FM line
is always continuous. The reason for the latter fact is that the
critical exponents of the magnetization (v.e.v. of the Higgs field)
in this approximation is grater than $1/2$. The slope of the
FM-PM--FI-AM line generally changes in the point $B$. However the
magnitude of this change depends strongly on the number of the
fermion flavours. We show that for given system (form of the lattice
fermion action) there exists such number $n_0$ (generally
noninteger), that for the number of the fermion flavours $n_f = n_0$
the slope of this line is continuous, too. The qualitative picture
for the phase diagrams is shown in Fig. 1. By a product, we find that
in the FI phase in the vicinity of the point $B$ the magnetization
and the staggered magnetization align orthogonally to each
other in the group space.

We want to emphasize that all the results of \cite{TZ} concerning the
phase structure of the systems in the regions bordering on the PM
phase are not affected. The slopes of the critical lines in those
regions are determined by the terms in the free energy quadratic in
the mean fields. Such terms has been calculated in the ladder
approximation developed in \cite{Ze,TZ}. For examination of the
region bordering on the FI phase one needs to know the free energy up
to the fourth order in the mean fields \cite{Zar}. In this case the
summation of the ladder diagrams becomes too complicated and the
approximation loses its advantages. At the same time, unjustified
neglecting the ladder diagrams may lead to obviously incorrect
results: for example, in the case of the SU(2) system with the naive
fermions the FI phase gets lost. Therefore we consider here the U(1)
systems for which the ladder diagrams do not contribute to the free
energy at least up to the terms quadratic in the mean fields. We
limit ourselves to the consideration of the weak coupling regime; the
examination of the strong coupling regime can be done easily in
analogous way.

\section{Free energy}

Our basic point is that the free energy for the U(1) systems in four
dimensions up to terms quartic in the mean fields has the form (all
the notations as in \cite{TZ})
\begin{eqnarray}
F = \lim_{N \rightarrow \infty} N^{-4} F_{MF}
&=&
- \frac{1}{2} a \, h^2
- \frac{1}{2} b \, h^{2}_{st}
+ \frac{1}{4} c' \, h^2 h^{2}_{st}
+ \frac{1}{4} c^{s} \, (h \cdot h_{st})^2 \nonumber \\
& & \mbox{} +\frac{1}{4} c^{v} \, (h \times h_{st})^2
+ \frac{1}{24} d \, h^4
+ \frac{1}{24} e \, h^{4}_{st}
+ \frac{1}{24} f \, h^4 \ln h^2,
\end{eqnarray}
where coefficients $a$ to $f$ are functions of $y$ and $\kappa$, and
$h \cdot h_{st} = \sum_a h^a h^{a}_{st}$, $h \times h_{st} = \sum_{a,
b} \epsilon^{a b} h^a h^{a}_{st}$.

The coefficients $a$ to $c$ and $e$ are calculated directly from the
eqs. (8)--(15) of \cite{TZ} and read as\footnote{We do not
take into account the correlations of the mean fields in the
quartic terms.}
\begin{eqnarray}
a(\kappa, y)&=&- \frac{1}{2} +  4 \kappa + y^2 n_f \int_p K^{2}(p),
\nonumber \\
b(\kappa, y)&=&- \frac{1}{2} -  4 \kappa + y^2 n_f
\int_p K(p) \cdot K(p+\pi), \nonumber \\
c'(\kappa, y)&=&- \frac{3}{8}
+ \frac{1}{2} y^2 n_f \int_p \left[K^{2}(p) + K(p) \cdot K(p+\pi)\right]
+ y^4 n_f \int_p K^{2}(p) K(p) \cdot K(p+\pi), \nonumber \\
c^{s}(\kappa, y)&=&- \frac{3}{4}
+ y^2 n_f \int_p \left[K^{2}(p) + K(p) \cdot K(p+\pi)\right]
+ \frac{1}{2} y^4 n_f \int_p K^{2}(p) K^{2}(p+\pi), \nonumber \\
c^{v}(\kappa, y)&=&- \frac{1}{2} y^4 n_f \int_p K^{2}(p) K^{2}(p+\pi),
\nonumber \\
e(\kappa, y)&=&- \frac{9}{8} - 12 \kappa +
3 y^2 n_f \int_p K(p) \cdot K(p+\pi)
\nonumber \\
& & + \frac{3}{2} y^4 n_f \int_p \left[2 \left(K(p) \cdot
K(p+\pi)\right)^2 - K^{2}(p) K^{2}(p+\pi)\right],
\end{eqnarray}
where $K_{\mu}(p) = L_{\mu}(p)/ L^2(p)$, so that $\int_p K^2(p)
\equiv G(0)$, $\int_p K(p) \cdot K(p+\pi) \equiv G(\pi)$ \cite{TZ}
(we omit the superscripts $W$ meaning the weak coupling regime).

The coefficients $f$ and $d$ in (1) cannot be calculated by the
term-wise integration of the standard weak coupling expansion of the
logarithm of the fermion determinant: the integrals of the $O(h^n)$
terms beginning from $n \geq 4$ have infrared divergences, since
$K^2(p) \rightarrow 1/p^2$ at $p^2 \rightarrow 0$. This is the reason
for appearing the logarithmic term in (1). It it this term that
differs our results from those of ref. \cite{Zar}.

To estimate the coefficient $f$, note that the contribution of the
fermion determinant to the free energy at $h_{st} = 0$ is given by
the expression
\begin{equation}
I = -2 n_f \int_p \ln \left[1 + y^2 \langle \phi \rangle^2 K^2(p)
\right].
\end{equation}
The integral is finite and certainly depends on the form of the
function $K(p)$. The point, however, is that the term $\propto h^4
\ln h^2$ comes from the infrared region $p^2 \sim 0$ which is not
sensitive to the detailed form of the function $K(p)$. Therefore, to
find the coefficient $f$ we can consider the integral with the same
infrared properties
\begin{eqnarray}
I' &=& - \frac{1}{(2 \pi)^2} n_f \int^{\Lambda}_{0} dp \:  p^3 \ln
\left( 1 + y^2 \langle \phi \rangle^2 \frac{1}{p^2} \right) \nonumber \\
&=& - \frac{1}{(4 \pi)^2} y^4 n_f \langle \phi \rangle^4 \ln \langle
\phi \rangle^2 + (\mbox{terms polynomial
in} \; \;  \langle \phi \rangle),
\end{eqnarray}
where $\Lambda = O(2 \pi)$ is an ultraviolet cutoff. From (4), taking
into account the relation $\langle \phi \rangle^2 = h^2/4 + O(h^4)$,
we find
\begin{equation}
f(\kappa, y) = - \frac{3}{32 \pi^2} y^4 n_f.
\end{equation}

Since we are interested in the domain of the phase diagram where
both $h$ and $h_{st}$ tend to zero, and therefore $|h^4 \ln h^2| \gg
h^4$, one can neglect the term $\propto d \, h^4$ in (1).

The FM-PM and AM-PM critical lines are determined by the equations
$a(\kappa, y) = 0$ and $b(\kappa, y) = 0$, respectively, so that the
coordinates of the point $B$ are
\begin{equation}
\kappa_B = - \frac{G(0) - G(\pi)}{8 [G(0) + G(\pi)]},
\quad
y^{2}_{B} = \frac{1}{n_f [G(0) + G(\pi)]}.
\end{equation}

In the vicinity of the point $B$ we have (in the notations of
\cite{Zar})
\begin{eqnarray}
&&a(\kappa, y) = a_{\kappa} (\kappa - \kappa_B) + a_y (y - y_B) +
O(\Delta^2), \nonumber \\
&&b(\kappa, y) = - b_{\kappa} (\kappa - \kappa_B) - b_y (y - y_B) +
O(\Delta^2), \nonumber \\
&&c'^{_, s_, v}(\kappa, y) = c'^{_, s_, v}_{B} +
O(\Delta), \quad
e(\kappa, y) = e_{B} + O(\Delta),
\quad
f(\kappa, y) = f_{B} + O(\Delta),
\end{eqnarray}
with
\begin{equation}
a_{\kappa} = 4, \quad
a_y = 2 n_f G(0) y_B, \quad
b_{\kappa} = 4, \quad
b_y = -2 n_f G(\pi) y_B,
\end{equation}
and $\Delta = y - y_B$ or $\kappa -\kappa_B$.

{}From (2) and (5) it follows that $c^{s}_{B} > 0$ and $c^{v}_{B} < 0$.
Hence, in the FI phase (if it exists) the mean fields $h$ and
$h_{st}$ align in such a way that $h \cdot h_{st} = 0$ and $h
\times h_{st} \neq 0$, i.e. orthogonally to each other in the group
space. Therefore, in the vicinity of the point $B$ the expression (1)
for the free energy is reduced to
\begin{eqnarray}
F &=& - \frac{1}{2} [a_{\kappa} (\kappa - \kappa_B) + a_y (y - y_B)]
 \, h^2 + \frac{1}{2} [b_{\kappa} (\kappa - \kappa_B) + b_y (y - y_B)]
 \, h^{2}_{st} \nonumber \\
&& \mbox{}+
\frac{1}{4} c_B \, h^2 h^{2}_{st} + \frac{1}{24} e_B \, h^{4}_{st} +
\frac{1}{24} f_B \, h^4 \ln h^2 + o(\Delta^2),
\end{eqnarray}
where $c_B = c'_{B} + c^{v}_{B}$. From (9) necessary
conditions for the existence of the FI phase follow:
\begin{equation}
c^{2}_{B} < \frac{1}{9} e_B f_B  \ln h^2, \quad e_B > 0, \quad f_B <
0
\end{equation}
(provided $a > 0$, $b > 0$). Hence, quite near the point B, where $h
\rightarrow 0$, the FI phase can exist in fact at any value of the
coefficient $c_B$. From (5) it follows that the third condition in
(10) is satisfied always, and we shall demonstrate in the next
section that the second condition for the systems considered in
\cite{TZ} is satisfied, too.

\section{Slopes of the critical lines and the FI phase}

Now, following the consideration of ref. \cite{Zar} setting up the
relation between the slopes of the critical lines and the critical
exponents of the magnetizations, one can easily find the slopes at
the point $B$.

Consider first the $h = 0$ critical line. Near the point $B$ it is
determined by the equation
\begin{equation}
- a(\kappa, y) + \frac{1}{2} c_B h^{2}_{st} = 0.
\end{equation}
{}From (9) it follows that nonzero value of the staggered magnetization
is determined by the equation
\begin{equation}
h^{2}_{st} = - \frac{6}{e_B} \left[b_{\kappa}(\kappa - \kappa_B)
+ b_y (y - y_B)\right].
\end{equation}
Then, from (11) and (12), in full correspondence with the results of
ref. \cite{Zar}, we have
\begin{eqnarray}
\left. \frac{d \kappa}{d y}\right|^{FM-PM}_{B}
&=& - \left. \frac{\partial a(\kappa, y) /
\partial y}
{\partial a(\kappa, y) / \partial \kappa}\right|_{B}
= - \frac{a_y}{a_{\kappa}}, \\
\left. \frac{d \kappa}{d y}\right|^{FI-AM}_{B}&=&-
\left.
\frac{\partial a(\kappa, y) /
\partial y
- (1/2) c_B \partial h^{2}_{st} / \partial y}
{\partial a(\kappa, y) / \partial \kappa
- (1/2) c_B \partial h^{2}_{st} /
\partial \kappa}\right|_{B} \nonumber \\
&=&- \frac{a_y + 3 b_y c_B/e_B}
{a_{\kappa} + 3 b_{\kappa} c_B/e_B }.
\end{eqnarray}

The things are different in the case of the $h_{st} = 0$ critical
line.  Indeed, though this line is determined by the equation
similar to (11):
\begin{equation}
-b(\kappa, y) + \frac{1}{2} c_B h^{2} = 0,
\end{equation}
the nonzero value of the magnetization is determined by the equation
($|\ln h^2| \gg 1$)
\begin{equation}
h^2 \ln h^2 =
\frac{6}{f_B} \left[a_{\kappa}(\kappa - \kappa_B)
+ a_y (y - y_B)\right].
\end{equation}
{}From (16) it follows that $\partial h^2 / \partial \kappa |_B =
\partial h^2 / \partial y |_B = 0$, i.e. the critical exponents
of the magnetization is grater than $1/2$. Therefore we have in this
case
\begin{eqnarray}
\left. \frac{d \kappa}{d y}\right|^{AM-PM}_{B}
&=& - \left. \frac{\partial b(\kappa, y) /
\partial y}
{\partial b(\kappa, y) / \partial \kappa}\right|_{B} = -
\frac{b_y}{b_{\kappa}}, \\
\left. \frac{d \kappa}{d
y}\right|^{FI-FM}_{B}&=&-
\left. \frac{\partial b(\kappa, y) /
\partial y
- (1/2) c_B \partial h^{2} / \partial y}
{\partial b(\kappa, y) / \partial \kappa
- (1/2) c_B \partial h^{2}/
\partial \kappa}\right|_{B} = - \frac{b_y}{b_{\kappa}},
\end{eqnarray}
i.e. the slope of the $h_{st} = 0$ critical line does not change in
the point $B$.

It is seen now that the FI phase appears always, provided the
conditions (10) are fulfilled. The change of the slope of the $h = 0$
critical line in the point $B$ is determined mainly by the value of
$c_B$: if $c_B = 0$ the slope does not change; if $c_B > 0$ the
domain with the FI phase gets narrow and in the limit $c_B
\rightarrow \infty$ it shrinks; on the contrary, if $c_B < 0$ the
domain with the FI phase becomes wider and in the limit $c_B \rightarrow
-\infty$ it occupies near the point $B$ all the region between the
FI-FM and AM-PM lines (Fig. 1).

We shall demonstrate now that all the three variants: $c_B = 0$, $c_B
> 0$, and $c_B < 0$ are realised for the systems considered in
\cite{TZ}. The FM-PM and AM-PM lines intersect in three of those
systems: in the systems with the SLAC, Weyl, and the mirror fermion
actions. In these cases we have: $I_1 \equiv \int_p K^2(p)
K^2(p+\pi) \approx 8.23 \times 10^{-3}$, $I_2 \equiv \int_p
\left( K(p) \cdot K(p+\pi) \right)^2 \approx 3.73 \times 10^{-3}$,
$I_3 \equiv \int_p K^2(p) K(p) \cdot K(p+\pi) \approx -7.83 \times
10^{-3}$ for the SLAC action; $I_1 \approx 4.59 \times 10^{-4}$, $I_2
\approx 1.86 \times 10^{-4}$, $I_3 \approx -6.09 \times 10^{-4}$ for the
Weyl action; $I_1 \approx  5.92 \times 10^{-5}$, $I_2 \approx 5.92
\times 10^{-5}$, $I_3 \approx -2.70  \times 10^{-5}$ for the mirror
fermion action (the values of $G(0)$ and $G(\pi)$ are given in
\cite{TZ}).

Although $e_B$ depends on the number of flavours $n_f$, in all these
cases $e_B > 0$ for any $n_f$ and therefore the FI phase appears in
all of these systems. The sign and the value of $c_B$ however depends
strongly on $n_f$: this is due to the fact that the $O(y^{0}_{B})$
and $O(y^{2}_{B})$ terms in $c_B$ are independent of $n_f$, while the
$O(y^{4}_{B})$ terms are $\propto 1/n_f$.  Moreover, for given system
there exists such number $n_0$, that $c_B = 0$ if $n_f = n_0$, $c_B >
0$ if $n_f > n_0$, and $c_B < 0$ if $n_f < n_0$. For our systems we
find:
\begin{eqnarray}
n_0 &\approx& 32 \quad \mbox{for the SLAC action}, \nonumber \\
    &\approx& 4.7   \quad \mbox{for the Weyl action}, \nonumber \\
    &\approx& 1.3   \quad \mbox{for the mirror fermion action}.
\end{eqnarray}
The corresponding phase diagrams are shown qualitatively in Fig. 1.

\section{Summary}

Thus, we have demonstrated that in the U(1) Higgs-Yukawa systems with
the SLAC, Weyl, or mirror fermion actions the FI phase does appear in
the weak coupling regime in the point $B$. In the FI phase, at least
quite near this point, the magnetization $h$ and the staggered
magnetization $h_{st}$ align orthogonally to each other in the group
space. Due to the fact that the free energy involves the logarithmic
term, that, in turn, leads to the critical exponents of the
magnetization $h$ grater than $1/2$, the $h_{st} = 0$ critical line
(the AM-PM--FI-FM line) is continuous together with its first
derivative in the point $B$. The derivative of the $h = 0$ critical
line (FM-PM--FI-AM line) is generally discontinuous in the point $B$.
However the discontinuity depends strongly on the number of the
fermion flavours $n_f$, so that in each system there exist such
number $n_0$, that for $n_f = n_0$ the line is smooth, too.

To conclude, note that for the systems in which one has $K(p + \pi) =
- K(p)$ (as it is, for example, for the naive fermions), the
logarithmic terms in the free energy appear also for the staggered
magnetization. This means that the critical exponents for both $h$
and $h_{st}$ are grater than $1/2$. Therefore, in the cases when the
FM-PM and AM-PM lines intersect in such systems, the slopes of both
critical lines are continuous in the point $B$. This may occur, for
example, in the case of the SU(2) system with the naive fermions. We
admit, however, that to make definite conclusion on the existence of
the FI phase in the SU(2) systems a special investigation is
necessary.

\section{Acknowledgements}

I am grateful to S.~Tominaga for the numerical calculations of the
integrals $I_1$ to $I_4$ and to J.~L.~Alonso for a correspondence.
The work was partly supported by the Russian Basic Research Fund
under grant 95-02-03868a.

\vskip 3cm

\newpage

\noindent {\Large \bf Figure captions}
\vskip 1cm
Fig. 1. Qualitative picture of the phase diagrams: (1) for $n_f <
n_0$, (2) for $n_f = n_0$, (3) for $n_f > n_0$.

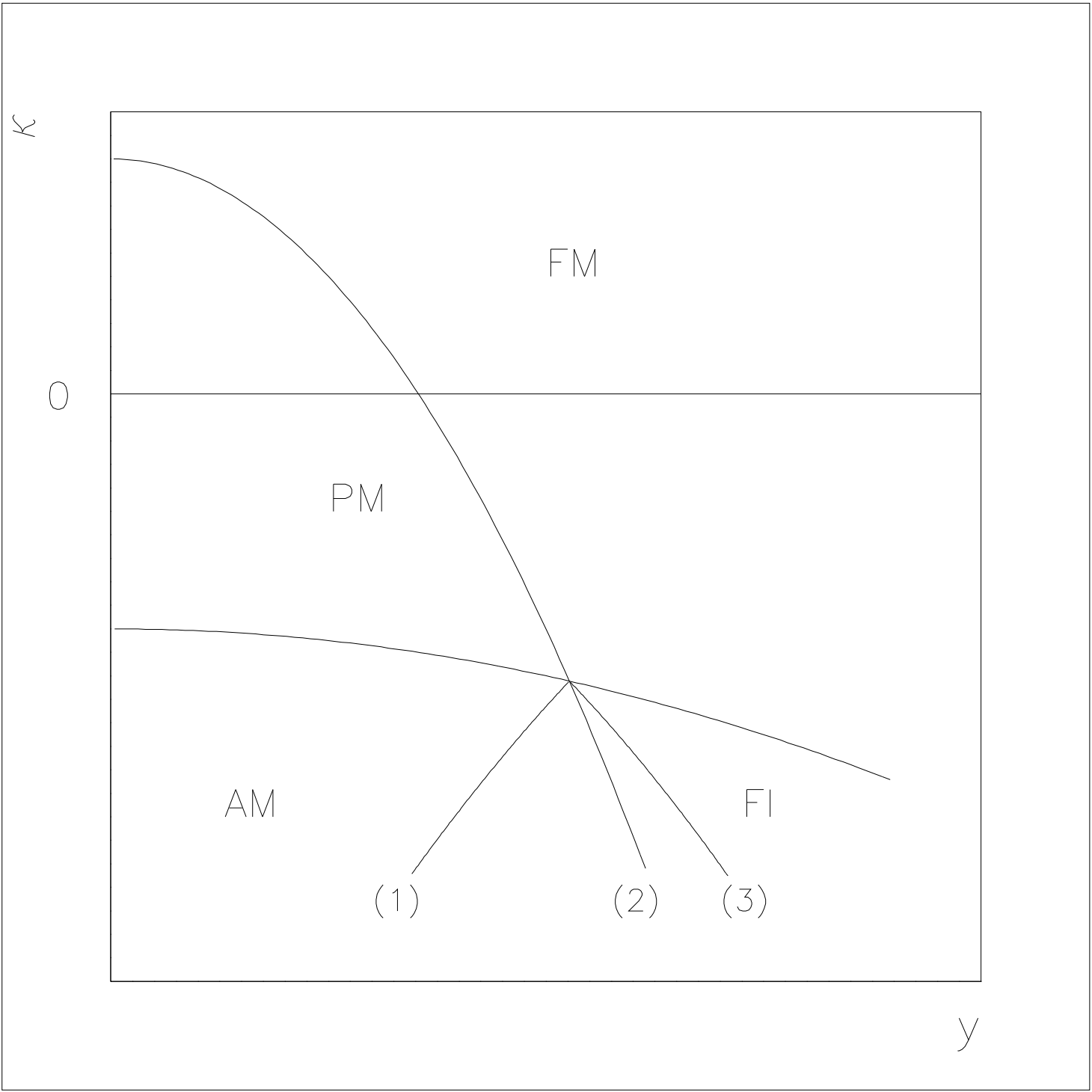
\end{document}